\documentclass[
aps,
twocolumn, 
showpacs,
prl, 
superscriptaddress] 
{revtex4-1}
\usepackage{graphicx}
\usepackage{color}
\usepackage{ulem}
\usepackage{verbatim}
\usepackage{amsmath}
\usepackage{amssymb}

\begin{document}

\title{Generation of single photons with highly tunable wave shape from a cold atomic quantum memory}%



\author{Pau Farrera}
\affiliation{ICFO-Institut de Ciencies Fotoniques, The Barcelona Institute of Science and Technology, 08860 Castelldefels (Barcelona), Spain}

\author{Georg Heinze}
\email[Contact: ]{georg.heinze@icfo.es}
\affiliation{ICFO-Institut de Ciencies Fotoniques, The Barcelona Institute of Science and Technology, 08860 Castelldefels (Barcelona), Spain}

\author{Boris Albrecht}
\altaffiliation{Present address: Niels Bohr Institute, University of Copenhagen, Denmark}
\affiliation{ICFO-Institut de Ciencies Fotoniques, The Barcelona Institute of Science and Technology, 08860 Castelldefels (Barcelona), Spain}

\author{Melvyn Ho}
\affiliation{Department of Physics, University of Basel, Klingelbergstrasse 82, 4056 Basel, Switzerland}

\author{Mat\'{i}as Ch\'{a}vez}
\affiliation{Department of Physics, University of Basel, Klingelbergstrasse 82, 4056 Basel, Switzerland}

\author{Colin Teo}
\altaffiliation{Present address: Singapore University of Technology and Design, 8 Somapah Road, 487372 Singapore}
\affiliation{Institute for Quantum Optics and Quantum Information of the Austrian Academy of Sciences, A-6020 Innsbruck, Austria}
\affiliation{Institute for Theoretical Physics, University of Innsbruck, A-6020 Innsbruck, Austria}

\author{Nicolas Sangouard}
\homepage{https://qotg.physik.unibas.ch}
\affiliation{Department of Physics, University of Basel, Klingelbergstrasse 82, 4056 Basel, Switzerland}

\author{Hugues de Riedmatten}
\homepage{http://qpsa.icfo.es}
\affiliation{ICFO-Institut de Ciencies Fotoniques, The Barcelona Institute of Science and Technology, 08860 Castelldefels (Barcelona), Spain}
\affiliation{ICREA-Instituci\'{o} Catalana de Recerca i Estudis Avan\c cats, 08015 Barcelona, Spain}%

\date{\today}

\begin{abstract}
We report on a single photon source with highly tunable photon shape based on a cold ensemble of Rubidium atoms. We follow the DLCZ scheme to implement an emissive quantum memory, which can be operated as a photon pair source with controllable delay. We find that the temporal wave shape of the emitted read photon can be precisely controlled by changing the shape of the driving read pulse. We generate photons with temporal durations varying over three orders of magnitude up to $10\,\mu\mathrm{s}$ without a significant change of the read-out efficiency. We prove the non-classicality of the emitted photons by measuring their antibunching, showing near single photon behavior at low excitation probabilities. We also show that the photons are emitted in a pure state by measuring unconditional autocorrelation functions. Finally, to demonstrate the usability of the source for realistic applications, we create ultra-long single photons with a rising exponential or doubly peaked wave shape which are important for several quantum information tasks. 
\end{abstract} 

\maketitle

A vast range of experiments in quantum information science and technology rely on single photons as carriers of information \cite{Sangouard2012}. Single photon sources are thus key components and have been continuously improved over the past years \cite{Eisaman2011}. The spectrum and temporal shape of the emitted photons are important parameters of such sources \cite{Raymer2012}. The generation of ultra-long single photons is for example an essential requirement for precise interactions with media exhibiting a sharp energy structure like trapped atoms, ions, or doped solids, which have been proposed as quantum memories for light \cite{Simon2010,Bussieres2013,Afzelius2015} and also with cavity optomechanical systems \cite{Bose1999,Marshall2003,Sekatski2014,Ghobadi2014,Aspelmeyer2014}. Several approaches to achieve narrow linewidth photons have been investigated, including e.g. cavity-enhanced spontaneous parametric down-conversion \cite{Bao2008,Haase2009,Fekete2013},  cold atomic ensembles \cite{Chou2004,Laurat2006,Thompson2006,Matsukevich2006a,Chen2006a,Du2008,Zhao2014}, single atoms \cite{McKeever2004, Hijlkema2007}, quantum dots \cite{Matthiesen2012} or trapped ions \cite{Almendros2009, Stute2012}. Moreover, significant efforts have been devoted to generate single photons with tunable temporal shapes \cite{Du2008,Almendros2009,Eisaman2004,Keller2004,Balic2005,Nisbet-Jones2011,Bao2012,Matthiesen2013,Zhao2015} which is of importance for many applications in quantum information science \cite{Aljunid2013,Liu2014}.

However, most of the previous approaches offered only a limited tuning range of the photon duration up to at most one order of magnitude \cite{Du2008,Almendros2009,Bao2012}. In this paper, we demonstrate a single photon source based on a cold $^{87}\mathrm{Rb}$ DLCZ-type \cite{Duan2001} quantum memory (QM) with a tuning range of three orders of magnitude, up to single photon durations of $10\,\mu\mathrm{s}$. Additionally, our QM allows us to release the single photons on demand after a programmable delay, which is essential for temporal synchronization tasks in quantum communication protocols as for example needed for quantum repeater architectures \cite{Briegel1998,Sangouard2011} or synchronization of photon pair sources \cite{Nunn2013}. We characterize the emitted photons by measuring their heralded and unheralded autocorrelation functions, demonstrating a high degree of anti-bunching and purity of the single photons. We finally demonstrate that ultra-long single photons with very flexible wave shapes are producible.

\begin{figure}
\includegraphics[width=.48\textwidth]{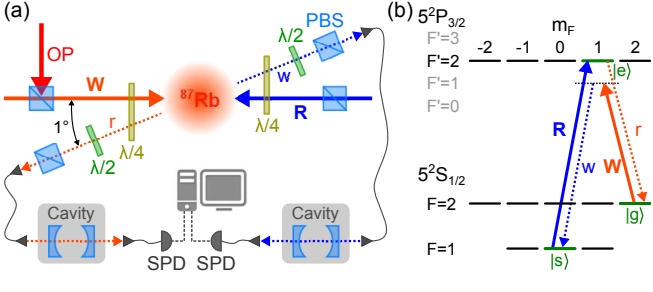}
\caption{(color online) (a) Experimental setup. Write pulse (W) and Read pulse (R) (non-dashed orange and blue arrows) are sent counter-propagating into the atomic cloud. Write and read photonic modes (w,r) are depicted by dashed blue and orange arrows. (b) Energy levels and coupling scheme for the DLCZ experiment. The color and line styles of the arrows correspond to the ones in (a).}
\label{Figure1}
\end{figure}

Our QM is based on a cold ensemble of $N$ identical $^{87}\mathrm{Rb}$ atoms in a magneto optical trap. Each atom exhibits a $\Lambda$-type level scheme consisting of a ground state $|g\rangle = |5^2S_{1/2},F=2,m_F=2\rangle$ a storage state $|s\rangle = |5^2S_{1/2},F=1,m_F=0\rangle$ and an excited state $|e\rangle = |5^2P_{3/2},F=2,m_F=1\rangle$ (see Fig.~\ref{Figure1}(b)). The atoms are initially prepared in the ground state $|g\rangle$ by optical pumping. A weak write pulse, $40\,\mathrm{MHz}$ red-detuned from the $|g\rangle\rightarrow|e\rangle$ transition, probabilistically creates a delocalized single-collective spin excitation (spin-wave) in the memory by transferring a single atom into the $|s\rangle$ state. This process is heralded by a Raman scattered write photon. The state of the spin-wave is to first order given by
\begin{equation}
\left|1_s\right\rangle=
\frac{1}{\sqrt{N}} \sum_{j=1}^{N}
e^{i\textbf{x}_j\cdot(\textbf{k}_W - \textbf{k}_w )} \left|g_1
\ldots s_j \ldots g_N\right\rangle, 
\label{eq1}
\end{equation}
where $\textbf{x}_j$ denotes the spatial position of the $j^{\mathrm{th}}$ atom and $\textbf{k}_W$ and $\textbf{k}_w$ are the wave vectors of the write pulse and the write photon respectively. Neglecting noise, the joint state of the write photon and the associated spin-wave is described by a two-mode squeezed state as 
\begin{equation}
\left|\phi\right\rangle = \sqrt{1-p}(\left|0_w\right\rangle \left|0_s\right\rangle + \sqrt{p}\left|1_w\right\rangle \left|1_s\right\rangle + p\left|2_w\right\rangle \left|2_s\right\rangle + o(p^{3/2})),
\label{eq2}
\end{equation}
with $p$ the probability to create a spin-wave correlated with a write photon in the detection mode. After a programmable delay, the spin-wave is converted back to a single read photon by a read pulse which is resonant with the $|s\rangle\rightarrow|e\rangle$ transition. Due to collective interference of all atoms, the read photon is emitted in a well defined spatial mode given by the phase matching condition $\textbf{k}_r = \textbf{k}_R + \textbf{k}_W - \textbf{k}_w$, where $\textbf{k}_R$ and $\textbf{k}_r$ are the wave vectors of the read pulse and read photon respectively. The raw retrieval efficiency is defined as $\eta_\mathrm{ret}=(p_{w,r}-p_{w,nr})/p_w$, where $p_{w,r}$ is the probability to detect a coincidence between a write and a read photon, $p_{w,nr}$ is the probability to detect a coincidence due to background noise and $p_w$ is the probability to detect a write photon per trial.

\begin{figure}
\includegraphics[width=.48\textwidth]{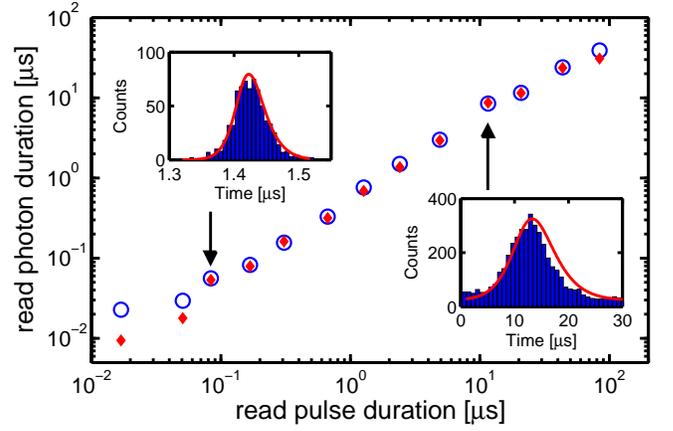}
\caption{(color online) Temporal duration (FWHM) of the read photon vs the duration of the driving read pulse. Experimental data (blue circles) are compared with numerical simulations (red diamonds). Error bars are smaller than the size of the symbols. The insets show two examples of the read photon wave shape as reconstructed from the number of counts and arrival times in the SPDs (blue histograms) as well as the simulated wave shapes (red lines) for which we allowed at most $10\%$ adjustment of the input parameters to account for experimental inaccuracies.}
\label{Figure2}
\end{figure}

The experimental setup is shown in Fig.~\ref{Figure1}(a). All light beams are derived from diode lasers modulated by acousto-optic modulators to address the $D_2$ line of $^{87}\mathrm{Rb}$ at $780\,\mathrm{nm}$. We combine a magnetic gradient of $20\,\mathrm{G/cm}$ with cooling light (red detuned from the ${|F=2\rangle\rightarrow|F'=3\rangle}$ transition) and repumping light (resonant with the ${|F=1\rangle\rightarrow|F'=2\rangle}$ transition) to load $N\approx 10^8$ Rubidium atoms into the MOT. After a $1.6\,\mathrm{ms}$ long optical molasses phase, we prepare all population in the $|g\rangle$ Zeeman sublevel by applying repumping light and $\sigma^+$ polarized optical pumping (OP) light on the ${|F=2\rangle\rightarrow|F'=2\rangle}$ transition. The spin-wave is generated by sending a write pulse of $15\,\mathrm{ns}$ duration (full width at half maximum FWHM), which is red detuned by $40\,\mathrm{MHz}$ from the $|g\rangle\rightarrow|e\rangle$ transition. The heralding write photon is collected at an angle of $1^\circ$ with respect to the write/read pulse axis. By changing the intensity of the write pulse, we can adjust the probability $p_w$ to detect a write photon per trial. For the experiments presented in this paper, $p_w$ ranges between $0.25\%$ and $1\%$ depending on the measurement. The read pulse, counterpropagating with the write pulse, is resonant with the $|s\rangle\rightarrow|e\rangle$ transition and its temporal shape can be precisely controlled. The read photon is collected in the same spatial mode but opposite direction of the write photon. By measuring the transmission of classical light sent trough the photons axis and by comparison of experimental and theoretical data in Fig.~\ref{Figure2} and Fig.~\ref{Figure5}, we consistently infer a coupling efficiency of the read photon into the first fiber of approximately $60\%$. The polarization of the write and read pulses in the frame of the atoms is $\sigma^-$ and $\sigma^+$ respectively, while the detected write and read photons are $\sigma^+$ and $\sigma^-$ polarized. We use a combination of quarter- and half-waveplates with polarization beamsplitters to transmit only the photons with the correct polarizations. The write and read photons are moreover spectrally filtered by identical monolithic Fabry-Perot cavities with approximately $20\%$ total transmission (including cavity transmission and subsequent fiber coupling), before being detected by single photon detectors (SPDs) with $43\%$ efficiency.

We now present the experimental results and compare them to detailed theoretical calculations. To generate read photons of variable length, we change the duration of the Gaussian-shaped read pulse as well as the storage time over several orders of magnitude (see Fig.~\ref{Figure2}). The shortest read pulse duration of approximately $17\,\mathrm{ns}$ leads to a read photon of around $23\,\mathrm{ns}$ duration. After that initial data point, we observe a quite linear increase of the read photon duration with the read pulse duration up to several tens of microseconds. The lower limit of photon duration is given by the limited optical depth $OD=5.5$ in our experiment which leads to limited superradiant emission of the read photon \cite{DeOliveira2014}, i.e. a much faster emission than the decay time of the excited state of approximately $27\,\mathrm{ns}$. A further technical limitation is given by the finite bandwidth of the spectral filtering cavity of about $60\,\mathrm{MHz}$ which additionally increases the detected duration for short read photons. This effect, together with the deviation from the adiabatic condition, partly explains the slight difference of the first data points in Fig.~\ref{Figure2} from the theoretical prediction (see below). In contrast, the upper limit of photon duration is given by the spin-wave linewidth which is mainly determined by thermal motion of the atoms and spurious external magnetic fields. This currently limits the maximal storage time in the memory of about $60\,\mu\mathrm{s}$, cf. supplemental material \cite{SM}. In addition to the spin-wave linewidth, the photon duration will also be limited by the coherence time of the read laser which has a specified linewidth of $20\,\mathrm{kHz}$. However, within the above limits we demonstrate that the photon duration is fully tunable and that the Gaussian wave shape of the driving read pulse is preserved in the read-out process (see insets). 

\begin{figure}
\includegraphics[width=.48\textwidth]{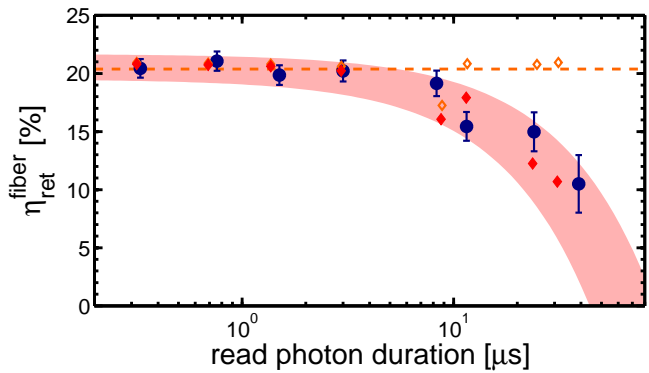}
\caption{(color online) Fiber-coupled retrieval efficiency $\eta_\mathrm{ret}^\mathrm{fiber}$ vs read photon duration (FWHM) for a write detection probability $p_w=0.5\%$. Experimental data (blue dots) are compared with numerical simulations for realistic (red diamonds) \cite{SM} and ideal (orange diamonds) conditions. The red shaded area depicts the expected range if the input parameters of the simulation are varied by $\pm10\%$.}
\label{Figure3}
\end{figure}

The dynamics of the write and read photon pairs is modeled using the Heisenberg-Lanvegin equations. For slowly varying optical fields propagating in a pencil-shape atomic ensemble, explicit expressions for both the write and read photon fields can be obtained in the adiabatic approximation \cite{Andre2005}. These field expressions can be subsequently used to reproduce the read photon emissions conditioned on the detection of a write photon from first and second order correlation functions, cf. supplemental material \cite{SM}. The result of these simulations which are based on independent measurements reproduce very well the experimental data presented in  Fig. \ref{Figure2}.

To characterize the retrieval efficiency of the photon source, we scanned the intensity of the driving read pulse for each duration. For short durations, we observe the expected Rabi oscillations in the retrieval efficiency vs. read pulse power \cite{Mendes2013}. If we generate photons with durations much longer than the natural decay time, the oscillations are damped and the efficiency approaches a constant value for high read pulse intensities \cite{SM}. The data of Fig.~\ref{Figure2} were taken with the read pulse intensity optimized for the highest possible efficiency. Fig.~\ref{Figure3} shows these optimized efficiencies vs the read photon duration. The plotted efficiency $\eta_\mathrm{ret}^\mathrm{fiber}$ corresponds to the probability of finding a read photon in the optical fiber after the vacuum cell, i.e. corrected for filtering and detector efficiencies only. We observe a constant retrieval efficiency of about $\eta_\mathrm{ret}^\mathrm{fiber}=20\,\%$ up to a read photon duration of approximately $10\,\mu\mathrm{s}$. Our numerical simulations match very well with the experimental data and also show that the efficiency in the constant region is just limited by the finite $OD$ of our atomic cloud. We verify numerically that in the absence of technical noise and considering infinite spin-wave coherence, for $OD=50$ an intrinsic retrieval efficiency of $80\%$ can be achieved while maintaining control of the photon shape. The later decrease of the efficiency at around $10\,\mu\mathrm{s}$ is due to dephasing of the spin-wave induced by atomic motion, spurious external magnetic field gradients \citep{Albrecht2015}, and to the finite read laser coherence time. In particular, our numerical simulations show clearly that in the absence of technical noise and in the limit of infinite spin coherence, the efficiency is kept constant (see orange diamonds and dashed line in Fig.~\ref{Figure3}).

\begin{figure}
\includegraphics[width=.48\textwidth]{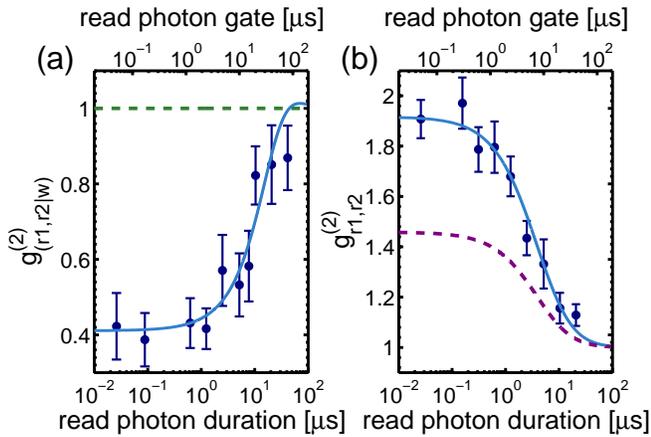}
\caption{(color online) Second order autocorrelation function of the generated read photons, (a) conditioned on the detection of a write photon in the same experimental trial at $p_w=0.25\%$ and (b) not conditioned on a write photon detection at $p_w=1\%$. The experimental data (blue dots) are compared with a theoretical model taking into account detector imperfections due to dark counts (blue lines). The dashed green line in (a) represents the classical bound of a coherent state and the dashed purple line in (b) shows the expected trace for a photon state with two modes.}
\label{Figure4}
\end{figure}

\begin{figure}
\includegraphics[width=.48\textwidth]{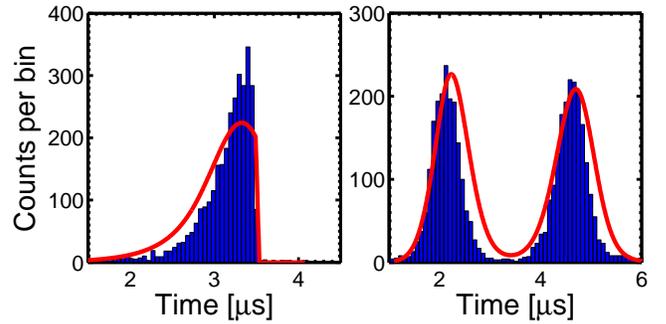}
\caption{(color online) Temporal wave shape of the read photon for two different driving wave shapes of the read pulse. Left, rising exponential and right a doubly peaked (time-bin) wave shape. Experimental data (blue histograms) are compared with numerical simulations (red line) for which we allowed at most $10\%$ adjustment of the input parameters with respect to the measured data. Both histograms were taken at $p_w=0.5\%$.}
\label{Figure5}
\end{figure}

Next, we characterized the state of the emitted read photons by measuring their heralded and unheralded second order autocorrelation functions depending on the read photon duration. To perform these measurements, we modified the setup and inserted a balanced fiber-beamsplitter into the read photon arm after the spectral filtering cavity, with both output ports connected to single photon detectors $r1$ and $r2$. First, we recorded the autocorrelation function conditioned on the detection of a write photon, defined as \cite{Grangier1986}:
\begin{equation}
g^{(2)}_{r1,r2|w}=\frac{p_{r1,r2|w}}{p_{r1|w}\cdot p_{r2|w}}
\label{eq3}
\end{equation}
where $p_{r1,r2|w}$ denotes the probability to measure a coincidence between both read photon detections conditioned on a write photon detection, and $p_{r1|w}$, $p_{r2|w}$ are the probabilities to detect a read photon via $r1$ or $r2$ conditioned on a write photon detection. The data shown in Fig.~\ref{Figure4}(a), clearly demonstrate the non-classicality of the photons (i.e. $g^{(2)}_{r1,r2|w}<1$) up to photon durations of more than $10\,\mu\mathrm{s}$. However, we don't reach the ideal value of $g^{(2)}_{r1,r2|w}=0$ of perfect single photons. For short read photon durations we are still limited by noise due to higher order components of the spin-wave which can be addressed by reducing the write probability $p_w$. In fact, the observed $g^{(2)}_{r1,r2|w}\approx 0.4$ is consistent with former measurements at similar values for $p_w$ and read pulse durations \cite{Albrecht2015}. For longer read photon durations we observe an increase of $g^{(2)}_{r1,r2|w}$ which can be simply explained by a higher number of dark counts of the SPDs for longer read photon detection gates (see upper axis in Fig.~\ref{Figure4}). The solid blue line shows the prediction of a non-perturbative theoretical model accounting for detector imperfections \cite{Sekatski2012}, for our measured dark count rate of $130\,\mathrm{Hz}$. The agreement between the model and the experimental data is excellent. 

The purity of the photon state is characterized by the unconditional autocorrelation function $g^{(2)}_{r,r}$ (see Fig.~\ref{Figure4}(b)). For an ideal two mode squeezed state, where the write and read photons are each emitted in a single temporal mode, one expects $g^{(2)}_{r,r}=2$ which is quite well fulfilled by the measured data up to a read photon duration of roughly $1\,\mu\mathrm{s}$. For longer durations, we observe a drop which can be attributed to either an increasing multimodality of the read photon ($g^{(2)}_{r,r}$ scales as $1+1/K$ with $K$ denoting the number of photon modes \cite{Christ2011}) or to measurement imperfections because of higher dark counts for larger detection gate widths. The solid blue line shows the prediction of the theoretical model accounting for experimental imperfections, for our measured dark count rate of $130\,\mathrm{Hz}$, assuming read photons emitted in a pure state. Our measurements follow very well this prediction, which suggests that the read photons are emitted mostly in a single mode. For comparison, we also plotted the expected behaviour for a single photon with $K=2$ modes (see purple dashed line) which significantly differs from the measured data, therefore confirming the single mode nature of the emitted read photons. Consequently, the read photons are close to be Fourier transform limited, giving linewidths ranging from around 20 MHz to less than 100 kHz.

Finally, we investigate the flexibility of the temporal wave shape of the generated read photons. Instead of a Gaussian shaped read pulse we send read pulses with a rising exponential envelope and a doubly peaked wave shape into the cloud. These two examples are of high importance for a broad class of applications in quantum information science and technology. Photons with rising exponential wave shape exhibit the highest possible absorbance when interacting with two-level systems \cite{Stobinska2009,Aljunid2013} and can be very efficiently loaded in optical cavities \cite{Bader2013,Liu2014}. The temporal shape of the generated rising exponential read photon is shown in the left plot of Fig.~\ref{Figure5}. The driving read pulse had a $1/e$ width of $300\,\mathrm{ns}$ and the data were taken at a write detection probability of $p_w=0.5\%$. We observe a similar retrieval efficiency of $\eta_\mathrm{ret}^\mathrm{fiber}=19.8\%$ as for a standard Gaussian shaped pulse of same duration (cf. Fig.~\ref{Figure3}). The conditioned autocorrelation function of the rising exponential photon is $g^{(2)}_{r1,r2|w}=0.31\pm 0.14$ (taken at $p_w=0.25\%$) and $g^{(2)}_{r1,r2|w}=0.73\pm 0.12$ (taken at $p_w=0.5\%$) which is clearly in the nonclassical regime. 

As a final example, we send a doubly peaked read pulse into the prepared QM. The intensity and duration of the first read-out peak was chosen such that the stored spin-wave is read out with half of the maximal efficiency and for the second peak the retrieval efficiency is maximized. This leads to a read photon with a shape shown in the right plot of Fig.~\ref{Figure5}. Photons with such a delocalized shape can be used to create time-bin qubits which have applications in robust long distance quantum communication \cite{Brendel1999,Marcikic2002}. The efficiency of the generated time-bin photon is $\eta_\mathrm{ret}^\mathrm{fiber}=25\%$, comparable to the standard Gaussian shaped photons, and the conditioned autocorrelation function is $g^{(2)}_{r1,r2|w}=0.54\pm 0.11$ (taken at $p_w=0.25\%$) and $g^{(2)}_{r1,r2|w}=0.75\pm 0.08$ (taken at $p_w=0.5\%$) which is clearly in the nonclassical regime. 

In conclusion, we demonstrated a highly flexible single photon source following the DLCZ protocol \cite{Duan2001} in a cold $^{87}\mathrm{Rb}$ atomic ensemble. Due to the storage capability of the source, we generated ultra-long single read photons which could be retrieved after a programmable delay after the heralding write photon. By varying the temporal width of the driving read pulse, the duration of the read photons could be changed over three orders of magnitude up to several tens of $\mu\mathrm{s}$. Up to a read photon duration of $10\,\mu\mathrm{s}$, we obtain a fiber-coupled retrieval efficiency of $\eta_\mathrm{ret}^\mathrm{fiber}=20\%$, which is just limited by the $OD$ in our experiment. We verified numerically that for $OD=50$ under ideal conditions, an intrinsic retrieval efficiency of $80\%$ can be achieved while maintaining control of the photon shape. The drop in retrieval efficiency at around $10\,\mu\mathrm{s}$ is mainly due to spin-wave dephasing induced by thermal motion, which is currently one of the main limitations in our setup. This could be improved by a more sophisticated trapping of the atoms \cite{Radnaev2010,Bao2012}. The generated read photons show a nonclassical behaviour up to durations of more than $10\,\mu\mathrm{s}$ for the heralded autocorrelation function and up to $1\,\mu\mathrm{s}$ we detect single photons in a pure state, currently just limited by the dark counts of our detectors. Finally, we demonstrate that our approach can be used to create single photons with a non-standard envelope like rising exponential or time-bin wave shapes, which have important applications in quantum information science and technology.

\section{Acknowledgements}
Research at ICFO is supported by the ERC starting grant QuLIMA, by the Spanish Ministry of Economy and Competitiveness (MINECO) and the Fondo Europeo de Desarrollo Regional (FEDER) through grant FIS2012-37569, by MINECO Severo Ochoa through grant SEV-2015-0522 and by AGAUR via 2014 SGR 1554. P.F. acknowledges the International PhD-fellowship program "la Caixa"-Severo Ochoa @ ICFO. G.H. acknowledges support by the ICFOnest+ international postdoctoral fellowship program. Research at the University of Basel is supported by the Swiss National Science Foundation (SNSF) through the Grant number PP00P2-150579 and the Army Research Laboratory Center for Distributed Quantum Information via the project SciNet. C.T. was supported by the Austrian Federal Ministry of Science, Research, and Economy (BMWFW) and would like to thank the hospitality of the quantum optics theory group in the University of Basel. \\

\bibliography{TunableSinglePhotons}
\bibliographystyle{prsty}

\onecolumngrid
\begin{appendix}
\vspace{30pt}
\begin{center}
\textbf{{\large Supplemental Material}}
\end{center}
\vspace{-10pt}
\newcommand{\ba}{\begin{eqnarray}}
\newcommand{\ea}{\end{eqnarray}}
\newcommand{\ban}{\begin{eqnarray*}}
\newcommand{\ean}{\end{eqnarray*}}
\newcommand{\moy}[1]{\langle #1 \rangle}
\newcommand{\ket}[1]{\mbox{$ | #1 \rangle $}}
\newcommand{\bra}[1]{\mbox{$ \langle #1 | $}}

\newcommand{\nonum}{\nonumber}

\section{PRINCIPLE OF NUMERICAL SIMULATION}\label{NumSim}
Here, we explain how one can compute the read photon properties conditioned on a write emission. 
We begin by solving the dynamics of the write field and the spin-wave. We then provide the explicit expression of the read field before showing how it can be used to obtain the conditional read photon characteristics.\\

\subsection{Write field and spin-wave expressions}
Working with a $\Lambda$-scheme for a three level system, we consider a writing pulse $\bar{\Omega}_W(t)$ detuned by $\Delta$ from the $| g \rangle \rightarrow| e\rangle $ transition. The $| e \rangle \rightarrow| s\rangle $ transition is characterized by an optical depth $\bar{d}_w$. $\gamma_{es}$ describes the decay of coherence in the $| e \rangle \rightarrow| s\rangle $ transition, and $\gamma_0$ describes the decay of the $| g \rangle \rightarrow| s\rangle $ coherence.
In the limit where the write field $\hat{\mathcal{E}}_w$  is slowly varying, propagating in a pencil-shaped atomic ensemble in which the $| g \rangle \rightarrow| e\rangle $ transition is driven by a off-resonant write pulse of duration $\tau_W$ satisfying $\gamma_{es} \tau_W \bar{d}_w \ll 1$, and also operating in the regime where $\Delta \gg |\bar{\Omega}_W|, \gamma_{es}$, the Raman scattering process results in the emission of a write field and the creation of a correlated spin-wave $\hat{S}$, whose dynamics are described with the Heisenberg-Langevin equations
\ba 
c \partial_{z'}  \hat{\mathcal{E}}_w &=& i \chi \hat{S}^\dagger \nonum \\
\partial_{t'} \hat{S}^\dagger &=&- \Gamma_S \hat{S}^\dagger - i \chi \hat{\mathcal{E}}_w + \hat{F}^\dagger _S 
\ea\
Here, we introduced shifted coordinates $z' = z$ and $t' = t - z/c$. $ \chi(t) =  (\sqrt{\bar{d}_w \gamma_{es} c / L}) \frac{\bar{\Omega}_W(t)}{\Delta} $, where $L$ is the length of the atomic medium, $\Gamma_S(t) = \gamma_S(t) + i \delta_S(t) $, $\gamma_S(t) = \gamma_0 + \gamma_{es} \frac{|\bar{\Omega}_W(t)|^2}{\Delta^2}$, $\delta_S(t) = - \frac{|\bar{\Omega}_W(t)|^2}{\Delta}$ and $\hat{F}_S$ is the Langevin noise operator for the write process.
The commutation relations for the relevant operators are given by
\ba
&& [\hat{\mathcal{E}}_w(z,t), \hat{\mathcal{E}}^\dagger_w(z',t')]  = L \delta[z - z' - c(t -t')] \nonum \\
&& [\hat{S}(z, t), \hat{S}^\dagger(z', t) ]= L \delta (z - z') \nonum \\
&& \langle \hat{F}_S(z,t) \hat{F}^\dagger_S(z',t') \rangle = 2 \gamma_S L \delta(z-z') \delta(t-t') \nonum \\
&& \langle \hat{F}^\dagger_S(z,t) \hat{F}_S(z',t') \rangle = 0 
\ea

The equations of motion can be solved as shown in Ref. \cite{Andre2005}. The solutions for the spin-wave and write field are
\ba \label{SpinWave}
\hat{S}^\dagger(z', t') &=& e^{-\Gamma(t')} \hat{S}^\dagger (z', 0)  \nonumber \\
&+& \int^{t'} _ 0 e^{-[\Gamma(t') - \Gamma(t'')]}\hat{F}_S ^\dagger (z', t'') dt'' \nonumber \\
&-&i \int ^{t'} _0 \chi(t'')e^{- [\Gamma(t') - \Gamma(t'')]} H(z', 0, t', t'')\hat{\mathcal{E}}_w (0, t'') dt'' \nonumber \\
&+& e^{-\Gamma(t')}\int^{z'} _0 G_s(z', z'', t', 0) \hat{S}^\dagger (z'',0) d z'' \nonumber \\
&+& \int ^{t'} _0 e^{-[\Gamma(t') - \Gamma(t'')]} \int^{z'} _0 G_s(z', z'', t', t'') \hat{F}_S ^\dagger (z'',t'') dz'' dt'' 
  \ea

and
  \ba \label{StokesField}
 \hat{ \mathcal{E}}_w (z', t') &=& \hat{\mathcal{E}}_w(0, t') \nonumber \\
  &+& i (\chi(t') / c) e^{- \Gamma(t')} \int ^{z'} _0 H(z', z'', t', 0) \hat{S}^\dagger (z'', 0) dz''  \nonumber \\
  &+& i (\chi(t') / c) \int ^{t'} _0 e^{- [\Gamma(t') - \Gamma(t'')]} \int ^{z'} _0 H(z', z'', t', t'') \hat{F}_S ^\dagger (z'', t'') dz'' dt'' \nonumber \\
  &+& (\chi(t') /c) \int ^{t'} _0 \chi(t'') e^{- [\Gamma(t') - \Gamma(t'')]} G_e(z', 0, t', t'') \hat{\mathcal{E}}_w(0, t'') dt''  ,
  \ea

where
\ba
H(z',z'',t',t'') &=& I_0 \left(2\sqrt{[g(t') - g(t'') ]\frac{z' - z''}{c}} \right) \nonum \\ 
G_s(z',z'',t',t'') &=&  \sqrt{\frac{g(t') - g(t'')}{c(z' - z'')}} I_1\left( 2\sqrt{[g(t') - g(t'') ]\frac{z' - z''}{c}} \right) \nonum \\
G_e(z',z'',t',t'') &=& \left( \frac{c(z' - z'')}{g(t') - g(t'')} \right) G_s(z',z'',t',t'')  \nonum .
\ea
Here, $I_{n}(x)$ refers to the modified Bessel function of the first kind, and here we have defined  $\Gamma(t) = \int_0^t \Gamma_S(t) dt$,  $g(t) = \int_0^t \chi(t')^2 dt' $. \\

\subsection{Read field expression}
During the retrieval process, a read pulse with Rabi frequency $\bar{\Omega}_R(t)$ is applied resonant with the $| e \rangle \rightarrow| s\rangle $ transition, converting the spin-wave in the atomic medium into a read field resonant with the $| e \rangle \rightarrow| g\rangle $ transition. The $| e \rangle \rightarrow| g\rangle $ transition is characterized by an optical depth $\bar{d}_{r}$. $\gamma_{eg}$ describes the decay of coherence in the $| e \rangle \rightarrow| g\rangle $ transition. 
Following similar arguments as Ref. \cite{Andre2005}, we can find the explicit expression of the read field $\hat{\mathcal{E}}_{r}$ as a function of the spin-wave resulting from the write process. Here, we consider a write emission at time $t_i$, and a non-zero read field at time $t_d$ after the write pulse ends. For the retrieval, the read field is emitted backwards, towards the z=0 position of the atomic medium. In the regime of $\bar{d}_{r} \gg 1 $ and sufficiently long read field duration $ \tau_{r} \gg \frac{1}{\gamma_{eg} \bar{d}_{r}}$,

\ba \label{AntiStokesField}
&&\hat{\mathcal{E}}_{r}(0,t = t_d + \xi) \nonum \\
&=& -\frac{\bar{\Omega}_R (t)}{g \sqrt{N}}   e^{-\gamma_0 t} 
\int_{c \Delta \tau(t,\xi)}^{L+c \Delta \tau(t,\xi)} \frac{1}{\sqrt{2 \pi} \Delta l(t,\xi)}
 \text{exp} \left[ - \frac{1}{2} \left( \frac{L-z}{\Delta l(t,\xi)} \right)^2\right] \hat{S}\left(L - z+c \Delta \tau(t,\xi),\xi \right) dz {} \nonum  \\
 &- & \frac{\bar{\Omega}_R(t)}{g\sqrt{N}} \int_\xi ^t \int_{c \Delta \tau(t,\xi))} ^ {L + c \Delta \tau(t,\xi)} \frac{e^{-\gamma_0 (t-t')}}{\sqrt{2 \pi} \Delta l (t, t')} \text{exp}\left[ -\frac{1}{2} \left( \frac{L-z}{\Delta l (t, t')} \right)^2 \right] \times \nonum \\
  && \Bigg[ \ \hat{F}_S(L - z+c \Delta \tau(t,t') ,t') {} \nonum \\
& & + \  i \frac{\Delta l^2(t,t') + (L-z)(2c(\Delta \tau(t,t')) + L - z)}{4 \gamma_{eg} c^2 (\Delta \tau(t,t'))^2 } \bar{\Omega}_R(t') \hat{F}_P(L - z+c \Delta \tau(t,t') ,t')\Bigg ] dz \nonum \\
 & +&\frac{ i }{g \sqrt{N}} e^{-\gamma_0 t} \hat{F}_P(0,t),
 \ea

where the commutation relations are
\ba
&& [\hat{\mathcal{E}}_{r}(z,t), \hat{\mathcal{E}}_{r}^\dagger(z',t')]  = L \delta[z - z' - c(t -t')] \nonum \\
&& \langle \hat{F}_P(z,t) \hat{F}^\dagger_P(z',t') \rangle = 2 \gamma_{eg} L \delta(z-z') \delta(t-t') \nonum \\
&& \langle \hat{F}^\dagger_P(z,t) \hat{F}_P(z',t') \rangle = 0 \nonum .
\ea

Here, $\Delta \tau(t, t') = \frac{L}{\bar{d}_{r} \gamma_{eg} c} \int_{t'}^t \bar{\Omega}_R^2(t'')  dt''$, and $\Delta l(t,t') = \sqrt{\frac{2 L c}{\bar{d}_{r}} \Delta \tau(t,t')}$. $g$ refers to the coupling constant between a single atom and a single read photon, and $N$ corresponds to the number of interacting atoms. This can be expressed as $g^2 N = \frac{\bar{d}_{r} \gamma_{eg} c }{L}$. $\xi$ indicates a suitable time after the write pulse has ended, and where the read pulse is considered to begin, so as to perform the numerical integration for the retrieval. $\hat{F}_P$ is the Langevin noise operator for the retrieve process.

\subsection{Conditional retrieval efficiency}

Equipped with the above expressions for the optical fields and spin-wave, we can compute the expectation of read photon emissions conditioned on the emission of a write photon from

\ba \label{condprob}
\eta_{r|w} = \frac{c}{L}\frac{ \int \int \langle \hat{\mathcal{E}}^\dagger_w (L,t_i) \hat{\mathcal{E}}^\dagger_{r} (0,t) \hat{\mathcal{E}}_{r}  (0,t) \hat{\mathcal{E}}_w (L,t_i) \rangle dt_i  dt }{\int \langle \hat{\mathcal{E}}^\dagger_w (L,t_i) \hat{\mathcal{E}}_w (L,t_i) \rangle dt_i }  .
 \ea
 
Evaluating the expression $ \langle \hat{\mathcal{E}}^\dagger_w (L,t_i) \hat{\mathcal{E}}^\dagger_{r} (0,t) \hat{\mathcal{E}}_{r}  (0,t) \hat{\mathcal{E}}_w (L,t_i) \rangle $ requires the expression in Eq. (\ref{StokesField}) and only the first term in Eq. (\ref{AntiStokesField}), which one in turn develops using Eq. (\ref{SpinWave}). A tedious but straightforward computation then results in 12 nonzero terms, of which 3 terms are 4-point noise correlators. Such 4-point noise correlations can be evaluated with use of Isserlis' theorem, which allows a decomposition into 2-point noise correlators for Gaussian random variables.

This gives, for example, 
\ba
\langle \hat{F}_S(z_1, t_1) \hat{F}_S^\dagger (z_2, t_2) \hat{F}_S(z_3, t_3) \hat{F}_S^\dagger (z_4, t_4) \rangle &=& \langle \hat{F}_S(z_1, t_1) \hat{F}_S^\dagger (z_2, t_2)\rangle \ \langle \hat{F}_S(z_3, t_3) \hat{F}_S^\dagger (z_4, t_4) \rangle \nonum \\
&+& \langle \hat{F}_S(z_1, t_1) \hat{F}_S(z_3, t_3)\rangle \ \langle \hat{F}_S^\dagger (z_2, t_2) \hat{F}_S^\dagger (z_4, t_4) \rangle \nonum \\
&+&\langle \hat{F}_S(z_1, t_1)\hat{F}_S^\dagger (z_4, t_4)\rangle \ \langle \hat{F}_S^\dagger (z_2, t_2) \hat{F}_S(z_3, t_3) \rangle,  \nonum 
\ea
where only the first term survives since the normal-ordered 2-point noise correlators are zero.

Finally, from the coupling efficiency of the read emission into the first fiber $\eta_\mathrm{fiber}$, we can reproduce the fiber-coupled conditional retrieval efficiency $\eta_\mathrm{ret}^\mathrm{fiber}$ using

\ba \label{CondRet}
\eta_\mathrm{ret}^\mathrm{fiber} =  \eta_{r|w} \eta_\mathrm{fiber},
\ea



valid in the low photon number regime ($\eta_{r|w} \ll 1$).

\subsection{Read photon shape}
The explicit expression of the fields also allows us to predict the temporal dependence of the conditional read emission, in a similar way as above. In particular, the conditional read photon flux inside the first fiber is given by


\ba
n_{r}^{cond} (t) = \frac{c}{L}\frac{  \int \langle \hat{\mathcal{E}}^\dagger_w (L,t_i) \hat{\mathcal{E}}^\dagger_{r} (0,t) \hat{\mathcal{E}}_{r}  (0,t) \hat{\mathcal{E}}_w (L,t_i) \rangle dt_i }{\int \langle \mathcal{E}^\dagger_w (L,t_i) \mathcal{E}_w (L,t_i) \rangle  dt_i }  \eta_\mathrm{fiber}.
\ea

\vspace{20pt}
\section{CHARACTERIZATION OF THE QUANTUM MEMORY}

In this section, we present additional experimental characterizations of the quantum memory and compare them to simulations based on the formalism introduced in the former section. Note first that for simulating the experimental data, we use the following conventions for the write (read) Rabi frequencies and optical depths: $\Omega_W= 2 \bar{\Omega}_W = \langle e  | {\bf{d}} \cdot {\bf{E}}_W | g \rangle$ ($\Omega_R= 2 \bar{\Omega}_R$) with $\bf{d}$ the dipole operator and $ {\bf{E}}_W$ the electric field amplitude of the write pulse and $d_w=  2 \bar{d}_w $ ($d_r=  2 \bar{d}_r$) such that the attenuation of the outgoing light intensity decreases as $I(L) = e^{-d_w (d_r)} I(0).$ \\

\begin{figure}[htbp]
\includegraphics[width=8 cm]{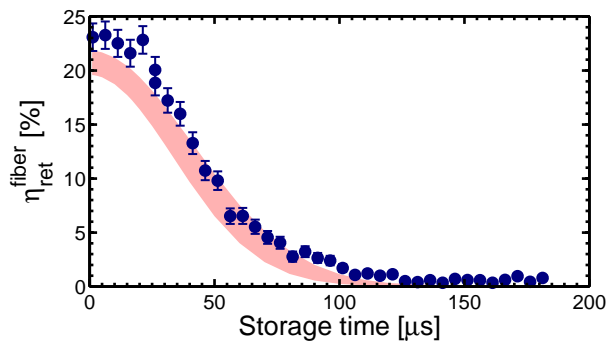}
\caption{Fiber-coupled conditional read efficiency $\eta_\mathrm{ret}^\mathrm{fiber}$ vs the delay between the write and read pulses. Experimental data (blue dots) are compared with numerical simulations (red shaded area). In order to account for the observed Gaussian decay reflecting an inhomogeneous broadening of the $| g \rangle \rightarrow| s\rangle $ transition, the exponential decay term $e^{-  \gamma_0 t}$ in Eq. (\ref{AntiStokesField}) is replaced with a Gaussian decay $e^{- \frac{1}{2} (t/{\gamma_0})^2}$, with $\gamma_0 = 53\,\mu\mathrm{s}$. The simulation is performed from Eq. (\ref{CondRet}) and uses the following measured values: The write pulse has a Rabi frequency of $\Omega_W=25.1\,\mathrm{MHz}$ and a FWHM of $15\,\mathrm{ns}$, detuned by $-40\,\mathrm{MHz}$ from the $| e \rangle \rightarrow| g\rangle $ transition. The peak Rabi frequency of the read pulse is at $23.5\,\mathrm{MHz}$ with a FWHM of $35\,\mathrm{ns}$. 
We take $d_w = 7.5$ and $d_r = 5$. An error of 10\% on the Rabi frequencies, pulse widths, optical depths and spin coherence time was assumed in order to obtain the bounds on the simulation.}
\label{fig1}
\end{figure}

\begin{figure}[htbp]
\includegraphics[width=8 cm]{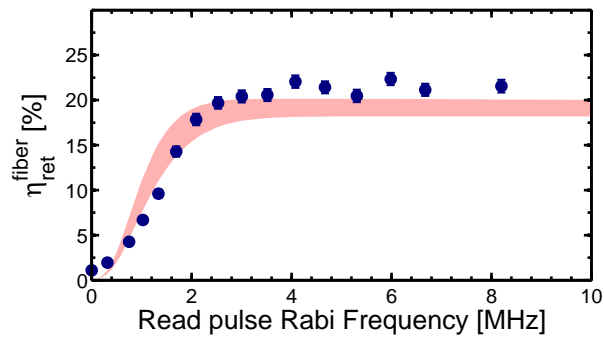}
\caption{Fiber-coupled conditional read efficiency $\eta_\mathrm{ret}^\mathrm{fiber}$ vs read Rabi frequency $\Omega_R$. Experimental data (blue dots) are compared with numerical simulations (red shaded area). The simulation is performed from Eq. (\ref{CondRet}) and uses the values of the write pulse presented in the caption of Fig. \ref{fig1}. The peak Rabi frequency of the read pulse is varied and its FWHM is $1.27\,\mu\mathrm{s}$. It has a delay of $2.16\,\mu\mathrm{s}$ from the write peak frequency. 
We take $d_w = 4.4$ and $d_r = 2.9$.  An error of 10\% on the Rabi frequencies, pulse widths, optical depths and spin coherence time was assumed in order to obtain the bounds on the simulation.}
\label{fig2}
\end{figure}

Fig. \ref{fig1} shows the fiber-coupled conditional retrieval efficiency as a function of the delay between the write and read pulses. To account for the observed Gaussian decay of the read efficiency, reflecting an inhomogeneous broadening of the $| g \rangle \rightarrow| s\rangle $ transition, all the simulations are performed by replacing the exponential decay term $e^{-  \gamma_0 t}$ in Eq. (\ref{AntiStokesField}) with a Gaussian decay $e^{- \frac{1}{2} (t/{\gamma_0})^2}.$ Fig. \ref{fig2} shows the conditional retrieval efficiency as a function of the delay between the write and read pulse. Note that the results of each Figure have been taken with slightly modified setups in each case, hence the differing optical depths. 

\vspace{10pt}
 
\newpage{}

\section{QUANTUM FEATURES AND PURITY OF THE READ PHOTONS}
 
To prove that the conditional read emission takes the form of single photons, we have measured the second order autocorrelation function conditioned on the detection of a write photon. Assuming that the write-read photon pairs are described by a two-mode squeezed vacuum state, an explicit expression of the conditional second order autocorrelation function can be derived in a non-perturbative way while taking the detector imperfections into account (non-unit, noisy and non-photon number resolving detectors), see Eqs. (24)-(25) in Ref. \cite{Sekatski2012}. The agreement between this model and the experimental data shows that the heralded second order autocorrelation function is mainly limited by dark counts, see Fig. 4(a) in the main text. \\

To demonstrate the purity of the read emission, we have also measured the (unconditional) second order autocorrelation function. Assuming again that the state of the write-read photon pairs corresponds to a two-mode squeezed vacuum, the exact expression of the second order autocorrelation function can be derived taking the detector imperfection into account, see formula $\tilde{g}_{\sum_n a_n}^{(2)}$ after Eq. (39) in Ref. \cite{Sekatski2012}. In particular, in the absence of noise and for small detection efficiencies, the auto-correlation function is given by $1+1/K,$ i.e. depends on the number of modes $K$. The full (blue) and dashed (purple) lines in Fig. 4 (b) of the main text are obtained by assuming that the read emission is pure and emitted in two possible modes respectively (with the detector imperfections). 

%
%
%

%
%
%
\end{appendix}

\end{document}